\documentclass[amsmath,amssymb,nofootinbib,twocolumn,reprint]{revtex4-1}
\pdfoutput=1

\usepackage[lmargin=50pt,rmargin=60pt,tmargin=60pt,bmargin=65pt]{geometry}
\usepackage{amstext,amsfonts,amsthm}
\usepackage[table]{xcolor} 
\usepackage{tikz}
\usetikzlibrary{positioning,arrows.meta,shapes,calc}
\usepackage{makecell}
\usepackage{braket} 
\usepackage{dsfont} 
\usepackage{graphicx} 
\usepackage[export]{adjustbox} 
\usepackage{subfig}
\usepackage{color}
\usepackage[colorlinks, linkcolor=blue, citecolor=blue, urlcolor=blue]{hyperref}

\theoremstyle{definition}

\theoremstyle{remark}

\newcommand{\mathtext}[2]{\texorpdfstring{#1}{#2}} 

\definecolor{cardinal}{rgb}{0.6,0,0}
\definecolor{darkgreen}{rgb}{0,0.5,0}
\definecolor{golden}{rgb}{0.92, 0.7, 0}
\definecolor{midnight}{rgb}{0, 0, 0.5}
\definecolor{darkblue}{rgb}{0.2, 0, 0.8}

\newcommand{\abs}[1]{\left| #1 \right|}

\newcommand{\be}{\begin{equation}}
\newcommand{\ee}{\end{equation}} 
\newcommand{\ba}{\begin{equation}\begin{aligned}}
\newcommand{\ea}{\end{aligned}\end{equation}}

\newcommand{\la}{\langle}
\newcommand{\ra}{\rangle}
\newcommand{\id}{\mathds{1}} 

\newcommand{\vphi}{\varphi}
\def\tend{\rightarrow}

\renewcommand{\(}{\left(}
\renewcommand{\)}{\right)}
\def\half{\frac{1}{2}}

\newcommand{\bigO}[1]{{O}\left(#1\right)}

\newcommand{\corr}[1]{\langle #1 \rangle}
\newcommand{\zcc}{\bar{z}}

\def \dd  {{\rm d}}

\newcommand{\Sgff}[1]{S_{\text{GFF}}[#1]}

\renewcommand{\a}{\alpha}

\newcommand{\eps}{\epsilon}

\renewcommand{\k}{\kappa}

\newcommand{\D}{\Delta}

\newcommand{\cF}{\mathcal{F}}

\newcommand{\cM}{\mathcal{M}}
\newcommand{\cN}{\mathcal{N}}
\newcommand{\cO}{\mathcal{O}}
\newcommand{\cP}{\mathcal{P}}
\newcommand{\cR}{\mathcal{R}}

\newcommand{\cT}{\mathcal{T}}

\newcommand{\cV}{\mathcal{V}}

\newcommand{\Z}{\mathbb{Z}}
\newcommand{\R}{\mathbb{R}}
\newcommand{\C}{\mathbb{C}}
\newcommand{\N}{\mathbb{N}}

\graphicspath{{fig/}}

\begin{document}
\title{The Lee-Yang model and its generalizations through the lens of long-range deformations}
\author{Fanny Eustachon}

\affiliation{CPHT, CNRS, École polytechnique, Institut Polytechnique de Paris, 91120 Palaiseau, France}

\begin{abstract}
        In two dimensions, the non-unitary class of conformal minimal models, $\cM(2,2m+1)$, has been recently conjectured to arise as renormalization-group fixed points of scalar field theories with complex $i\vphi^{2m-1}$ interaction, $m\in \N$, $m\ge2$. We test a variation of this conjecture through the perturbative study of two separate long-range constructions based on respectively the minimal model and its potential Landau-Ginzburg formalism. For $m>2$, inconsistencies are found when subsequently relating both constructions. In contrast, the long-range Lee-Yang model, the $m=2$ case, is shown to be analogue to the long-range Ising model.
\end{abstract}

\maketitle
      
\twocolumngrid
\renewcommand{\baselinestretch}{0.25}\normalsize
\tableofcontents
\renewcommand{\baselinestretch}{1.0}\normalsize

\section{Introduction}
\label{sec:introduction}

In two dimensions, using the power of representation theory, a classification of local conformal field theories (CFTs) has been determined, within which minimal models $\cM(p,q)$ belong. Those models are exactly solvable, allowing the exact computation of conformal data such as critical exponents and OPE coefficients.
In a quantum field theory (QFT) setting, while the enhancement of scale invariance to conformal invariance lacks a general proof, the converse holds. CFTs are intrinsically scale invariant by nature, and are expected to lie on infrared fixed points of the renormalization group flow. The corresponding UV theory is called a Landau-Ginzburg description.
Relatively few of them are known and even less understood. The prototypal example is the Ising multicritical class \cite{Zamolodchikov:1986db} $\cM(m,m+1)$ described by a Gaussian theory endowed with a $\vphi^{2(m-1)}$, $m\in\N$ interaction.
Following its spirit, a Landau-Ginzburg description for the Lee-Yang multicritical class $\cM(2,2m+1)$, $m\in\N$, $m\ge2$ has been recently proposed\footnote{A second proposal \cite{Lencses:2024wib} identifies the $i\vphi^{2m-1}$-theory to $\cM(2,4m-3)$. Our results will apply similarly to this conjecture.} \cite{Becker:1991nq,Becker:1991nr,vonGehlen:1994rp,Zambelli:2016cbw,Katsevich:2024jgq,Katsevich:2025ojk}: a free scalar bosonic theory deformed by a purely imaginary interaction $i\vphi^{2m-1}$. Its action then presents an antilinear ($\cP\cT$) symmetry $S^*[-\vphi]=S[\vphi]$, guarantying the spectrum's reality in absence of spontaneous symmetry breaking \cite{Gorbenko:2018ncu,Castro-Alvaredo:2017udm}.

However, while simple in nature, this class of theories is complicated to continue to $d=2$ for a number of reasons. First, in lower dimensions, those theories are strongly coupled and relatively far from their upper critical dimension, reducing the insight one can get from perturbation theory. Furthermore, the minimal models $\cM(2,2m+1)$ are non-unitary, meaning that they possess primary operators violating unitary bounds, have complex OPE coefficients, and thus violate assumptions behind the usual non-perturbative techniques such as the conformal bootstrap or Monte-Carlo techniques.
Nevertheless, a number of perturbative and non-perturbative studies have been successful in the special case of the Lee-Yang model ($m=2$) \cite{Codello:2017epp,Gracey:2025rnz,Borinsky:2021jdb,Gliozzi:2014jsa,ArguelloCruz:2025zuq,Fan:2025bhc,EliasMiro:2025msj,Zambelli:2016cbw,Benedetti:2026tpa}, but the conjecture for $m>2$ is still an open debate.

The purpose of this paper is to check the consistency of the conjecture through the lens of long-range models directly in $d=2$. 
The core idea will consist in building a long-range deformation of both the minimal model and its Landau-Ginzburg formalism and study their relationship using conformal perturbation theory. We will study separately the Lee-Yang model $\cM(2,5)$ and its multicritical class $\cM(2,2m+1)$.
Computational methods and constructions are closely related to the one used for long-range unitary minimal models. While most of the necessary arguments will be partially repeated, we refer to \cite{Behan:2025ydd} for technical details and more precise arguments.


\section{\mathtext{$2d$}{2d} Long-range Lee-Yang model}
\label{sec:LY}

{The Lee-Yang model can be interpreted as a statistical model describing the complex zeros of the Ising partition function, related to the Lee-Yang edge singularity. Its Landau-Ginzburg formalism, the $i\vphi^3$-theory, has been proposed by Fisher in 1978 \cite{Fisher:1978pf,Cardy:2023lha}. 
As a critical phenomenon, the Lee-Yang model is characterized by only one independent critical exponent, which translates in the CFT language into the minimal model $\cM(2,5)$ being composed of two Virasoro primary operators, the identity and the field $\phi_{1,2}$, of scaling dimension $\Delta_{1,2} = -4/5$.
In the following, we will consider two constructions of the long-range model based on either the $i\vphi^3$-theory or the short-range $\cM(2,5)$ and sketch their relations.
Long-range versions in $d=1$ of the Lee-Yang model have been mentioned in the literature \cite{ZGlumac_1991,Ghosh:2026gku}.
}
\medskip

{{\bf Long-range interactions.}
Nonlocal behaviour will be mostly implemented by introducing a generalized free scalar field (GFF) in $d$-dimension: 
\be\label{eq:GFF-S}
\Sgff{\vphi}=\frac{\cN}{2}\int \dd^d x_1\dd^d x_2 \frac{\(\vphi(x_1)-\vphi(x_2)\)^2}{\abs{x_1-x_2}^{d+s}},
\ee
such that the GFF scaling dimension\footnote{In general $d$, a local free field has a scaling dimension of $\Delta_{\vphi} = \frac{d-2}{2}$. } depends continuously on a real parameter $s\in\R$:
\be\label{eq:GFF-scaling}
\Delta_{\vphi} = \frac{d-s}{2}.
\ee
The normalization factor $\cN$ is choosen such that the two point function is normalized in position space $\corr{\vphi(x)\vphi(y)}_0 = \abs{x-y}^{-2\Delta_{\vphi}}$.
While such theory is nonlocal by design and thus noteworthily breaks any Virasoro symmetry or prevents the existence of a local stress energy tensor, the advantage of the long-range description lies in the presence of the extra tunable parameter $s$, independent of the spacetime dimension.}

After deformation of the action \eqref{eq:GFF-S} by an interaction term of the form $\int \dd^d x\, \cO(x)$, the theory can be regularized through the variation of the parameter $s$, setting $\Delta_{\cO}=d-\eps$, $\eps\ll1$ at a fixed value of $d$.
However, even in the presence of this term, the dimension of the fundamental field $\vphi$ is protected \cite{Fisher:1972zz} and sticks to its GFF scaling dimension \eqref{eq:GFF-scaling}, as the nonlocal kinetic term cannot be renormalized by local UV divergencies.\medskip


{\bf Deformation in $i\vphi^3$.}
Setting now $d=2$, let's consider the most natural construction of a long-range Lee-Yang model: the deformation of a GFF by a local purely imaginary interaction term:
\be\label{eq:MFT-S}
S_{\text{LR}}=S_{\text{GFF}}[\vphi]+\frac{i\lambda_0}{3!}\int \dd^2x\, \vphi^{3}(x).
\ee
The status of the interaction directly depends on the parameter $s$: at $s=\bar{s}$, the interaction is marginal with $3\Delta(\bar{s})= 2$; for $s< \bar{s} = 2/3$, the interaction is irrelevant, the model flows to a trivial fixed point, and thus belongs to the mean field theory regime; for $s>\bar{s}$, the interaction is relevant and the model flows instead to a non-trivial fixed point. The resulting fixed points forms a continuous family with respect to $s$, which is referred to as \textit{long-range Lee-Yang models}. \medskip

Near the mean field theory regime, the theory is weakly coupled, and can be determined perturbatively around $s = \bar{s} +\frac{2}{3}\eps$, $\eps\ll1$.
In a conformal perturbation theory setting \cite{Komargodski:2016auf,Behan:2025ydd}, the bare coupling $\lambda_0$ in \eqref{eq:MFT-S} is renormalized in the $\overline{\text{MS}}$ scheme by expanding the one-point function $\corr{\cO(\infty)e^{-i\lambda_0\int \dd^2x\, \cO(x)}}$ at weak coupling with a near marginal interaction operator $\Delta_{\cO}=2-\eps$. The resulting beta function for $\cO=i\vphi^3$ takes the form:
\be
\beta(\lambda)= -\eps \lambda +2\pi^2 \frac{\Gamma\(\frac{1}{3}\)^3}{\Gamma\(\frac{2}{3}\)^3}\lambda^3+\bigO{\lambda^5}.
\ee
The interaction being odd under a $\Z_2$-symmetry, all contribution from even order in $\lambda$ vanish identically.
Then, the fixed point equation $\beta(\lambda_*)=0$ presents three solutions: the trivial Gaussian solution and a complex conjugate pair of purely imaginary fixed points:
\be
i\lambda_* = \pm \frac{i}{\sqrt{2\pi^2}} \frac{\Gamma\(\frac{2}{3}\)^{\frac{3}{2}}}{\Gamma\(\frac{1}{3}\)^{\frac{3}{2}}}\sqrt{\eps}+\bigO{\eps}.
\ee

The anomalous dimension $\gamma_{\alpha}(\lambda_*)$ of a monomial operator $\vphi^\alpha$, $\alpha\in\N$, $\alpha>1$ at the fixed point is obtained in a similar fashion by renormalizing its two-point function, leading to an infrared scaling dimension given by $\Delta_{\vphi^\alpha}|_{\lambda=\lambda_*} =\alpha\Delta_{\vphi}+\gamma_{\alpha}(\lambda_*)$, or after computation,
\be\label{eq:MFT-ano}
\Delta_{\vphi^\alpha}|_{\lambda=\lambda_*} = \frac{2}{3}\alpha+\frac{\alpha(3\alpha-5)}{6}\eps +\bigO{\eps^2}.
\ee
For $\alpha=2$, the expression \eqref{eq:MFT-ano} can be recast as $\Delta_{\vphi^2}|_{\lambda=\lambda_*} = 2-\Delta_{\vphi}$. However, the operator $\vphi^2$ should not be thought as a descendant of the field $\vphi$, as done in the short-range theory, but instead as a physical shadow operator, ensuing from the nonlocal Schwinger-Dyson equations. Therefore, the anomalous dimension of $\vphi^2$ is exact at all order in perturbation theory.
\medskip

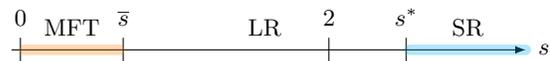
\begin{figure}[h]
   \begin{minipage}{\columnwidth}
    \centering
    \begin{tikzpicture}[glow/.style={%
    preaction={draw,line cap=round,line join=round,
    opacity=0.3,line width=4pt,#1}},glow/.default=yellow,
    transparency group]
      \tikzset{line/.style={thin}}
     
      \node[label=right:$s$] (a)[inner sep=0pt]  {};
      \node (d) [left =0.8\textwidth of a, inner sep=0pt] {};
      \path[draw,-latex] (d) to (a);

    \node[label={[yshift=1pt]$0$}] (tick1) [left=0.78\textwidth of a, inner sep=0pt] {$\vert$};
    \node[label={[yshift=1pt]$s^*$}] (tick2) [left=0.18\textwidth of a, inner sep=0pt] {$\vert$};
    \node[label={[yshift=1pt]$\overline{s}$}] (tick3) [left=0.62\textwidth of a, inner sep=0pt] {$\vert$};
    \node[label={[yshift=1pt]$2$}] (tick4) [left=0.3\textwidth of a, inner sep=0pt] {$\vert$};  
    \path[glow={cyan}] (a) -- (tick2) node [midway, above,yshift=2pt] {SR};
    \path (tick2) -- (tick3) node [midway, above,yshift=2pt] {LR};
    \path[glow={orange}] (tick3) -- (tick1) node [midway, above,yshift=2pt] {MFT};
      

  \end{tikzpicture}%
    \caption{\label{fig:phase}Schematic summary on the expected universality class of the IR fixed point depending on the value $s$.}
    \end{minipage} 
\end{figure}

As the value of $s$ increases, Sak \cite{Sak:1973oqx} theorized that, in the Ising model, the anomalous dimension of the dangerously irrelevant operator $\partial_{\mu}\vphi\partial^{\mu}\vphi$ increases with $s$ enough to transform it to a relevant operator at some critical $s=s^*$. For $s>s^*$, the IR fixed point then becomes unstable and flows instead to the short-range universality class. 
The spectrum continuity condition imposes that the crossover should happen when the protected scaling dimension $\Delta_{\vphi}$ hits 
the smallest conformal dimension with respect to the absolute value norm in the spectrum of the corresponding short-range minimal model.
Assuming a similar scenario for the Lee-Yang model, this crossover happens when $\Delta_{\vphi}(s^*)=\Delta_{1,2}$, that is at $s^* = 14/5 = 2.8$. The different regimes are summarized in figure \ref{fig:phase}.

Due to the negative sign of $\Delta_{1,2}$, the crossover is forced to happen at $s^*>2$. However, the limit $s\tend 2$ remains a puzzle. We elaborate on this issue in the discussion.\medskip

{\bf Deformation of $\cM(2,5)$.}
Let us now consider a second independent construction of a long-range minimal model: the deformation of the local short-range $\cM(2,5)$ by its weak coupling to a nonlocal GFF \footnote{Not to be confused with $\vphi$ in \eqref{eq:MFT-S}.} $\chi$:
\be\label{eq:SR-S}
S'_{\text{LR}}=S_{\cM(2,5)}+S_{\text{GFF}}[\chi]+g_0\int \dd^2x\, \phi_{1,2}\chi.
\ee
In the UV, the dimension of $\phi_{1,2}$ is fixed at its conformal value $\Delta_{1,2}$, while the scaling dimension of $\chi$ depends on $s$ through $\Delta_{\chi}=2-\Delta_{1,2}-\delta$, with $\delta=s^*-s\ll 1$. 
For $s\ge s^*$, the operator $\cO = \phi_{1,2}\chi$ is irrelevant, and the theory flows to a decoupled product of a GFF theory and the short-range model. Rather, for $s<s^*$, the theory is again expected to flow to a non-trivial fixed point, the long-range fixed point.
Then, in the IR, the Schwinger-Dyson equations imply now the following shadow relation:
\be
\Delta_{1,2}|_{g=g_*} = 2- \Delta_{\chi}.
\ee
It entails that the IR dimension of $\phi_{1,2}$ is controlled by the protected dimension of $\chi$, and its anomalous dimension is fixed correspondingly at $\gamma_{1,2} = \delta$ at all order in perturbation theory.\medskip

The above flow \eqref{eq:SR-S} is part of a more general construction and corresponds to the long-range minimal model of type $\text{LRMM}_{2,5}(1,2)$. This particular choice of coupling to $\phi_{1,2}$ shares the same underlying motivations as the unitary $\text{LRMM}_{m,m+1}(2,2)$ \cite{Behan:2025ydd}, which can be summarized as follows:
\begin{enumerate}
    \item From \eqref{eq:SR-S}, the cross-term in \eqref{eq:GFF-S} is retrieved for $g_0\in \R$ by integrating out $\chi$
    \be \label{eq:SR-intergatingout}
    S \supset -\half g_0^2 \int \dd^2 x_1\dd^2x_2\,\frac{\phi_{1,2}(x_1) \phi_{1,2}(x_2)}{\abs{x_1-x_2}^{2\Delta_{\chi}}},
    \ee
    with $\phi_{1,2}\sim \vphi$ acting as the order parameter. 
    The negative sign matches with the full nonlocal kinetic term \eqref{eq:GFF-S}, it is the one that ensures positivity of the action, once the singular terms at $x_1\sim x_2$ have been subtracted, leading to \eqref{eq:GFF-S};
    \item The identification $\phi_{1,2}\sim \vphi$ is further supported by the absence of wavefunction renormalization in both theories and the protected dimension of $\phi_{1,2}$ at $s < s^*$;
    \item Then, the extra field $\chi$, the shadow of $\phi_{1,2}$, is identified to the extra primary $\phi^2$, the shadow of $\vphi$ in \eqref{eq:MFT-S}, and therefore preserving the continuity of the spectrum. Consequently, in term of $s$, $\Delta_{\chi}=(2+s)/2$.
\end{enumerate}

However, the above conditions, while necessary, are not sufficient, and the duality between both long-range constructions still lacks a full proof, which would require non-perturbative computations between $\overline{s}$ and $s^*$, a study of the nonlocal currents, as well as a full understanding of the limit $s\tend 2$.
Thus, in the following, both constructions are to be assumed independent, with the aim to check the consistency of the above listed properties.\medskip

In the conformal perturbation setting, at $\Delta_{\cO} = 2 -\delta$, with $\cO=\phi_{1,2}\chi$, the beta function takes the general form, 
\be\label{eq:SR-beta}
\beta(g)=-\delta g + \beta_3 g^3 + \bigO{g^5},
\ee
where the absence of even contribution in $g$ are due to the linearity in $\chi$ of the interaction term $\cO$. 
The coefficient $\beta_3$ corresponds to the regularized integral over the domain
$\cR =\left\{z:\abs{z}<1, \abs{z}<\abs{z-1}\right\}$:
\be\label{eq:SR-beta3}
\beta_3 = - \pi \left.\int_{\cR}\dd^2 z\, \corr{\cO(0)\cO(z,\overline{z})\cO(1)\cO(\infty)}\right|_{\text{finite}}.
\ee
The four-point function is evaluated in the unperturbed theory, $\text{GFF}\times\cM(2,5)$, and is determined from the second order differential BPZ's equation \cite{DiFrancesco:1997nk}
\ba\label{eq:SR-4pts}
    &\la\cO(0) \cO(z, \bar{z}) \cO(1) \cO(\infty) \ra = \frac{1 + |z|^{-2\D_\chi} + |1 - z|^{-2\D_\chi}}{|z|^{2\D_{1,2}}}\times\\&\sum_{k \in \{ 1,3 \}} C_{(1,2)(1,2)(1,k)}^2 \left | \mathcal{F}_{(1,k)}^{(1,2)(1,2)(1,2)(1,2)}(z,\bar{z}) \right |^2\,,
\ea
where the OPE coefficients $C_{(1,2)(1,2)(1,r)}\in\C$ and blocks $\mathcal{F}_{(1,r)}^{(1,2)(1,2)(1,2)(1,2)}$ are known in a closed form, with the latter depending on hypergeometric functions (see \cite{Mussardo:2010mgq} or appendix).
Following the prescription proposed in \cite{Behan:2017emf,Behan:2025ydd}, $\beta_3$ can be evaluated numerically \footnote{Even considering the analytical control given by hypergeometric functions, the integral remains hard to compute analytically due to the complicate domain of integration.} to arbitrary precision:
\be
\beta_3 = (1.4019049631\pm5\cdot 10^{-10}) >0.
\ee
The beta function \eqref{eq:SR-beta} then admits two \textit{real} non-trivial fixed point solutions of the form $g_*^2 = \delta/\beta_3$, such that \eqref{eq:SR-intergatingout} is consistent with sign of the cross-term in \eqref{eq:GFF-S}.
In terms of IR scaling dimension, using the same prescription, the anomalous dimension $\gamma_{1,2} = \delta$ is verified numerically up to machine precision. The scaling dimension of the interaction term $\cO$ is obtained from the derivative of the beta function \eqref{eq:SR-beta} evaluated at the fixed point: $\Delta_\cO|_{g=g_*} = 2+2\delta+\bigO{\delta^2}$.\medskip

Lastly, spinning operators belonging to the Virasoro identity multiplet acquire an anomalous dimension in the IR through multiplet recombination  \cite{Rychkov:2015naa,Giombi:2016hkj,Behan:2017emf} arising from the broken conformal Ward identities $\overline{\partial}T^I=g_* b^I_J\cV^J+\bigO{g_*^2}$, where $b$ is a matrix of numerical coefficients. Assuming spectrum continuity, the operators $\cV^J$ should correspond to Virasoro primary operators of dimension $(h,\overline{h})=(\text{spin},1)$ in the unperturbed theory.
At spin\footnote{The Verma module of the identity operator admits a null state at both level 1 and 4 in $\cM(2,5)$, preventing the existence of a quasiprimary current at higher spin.} $2$, that is in the case of the stress-energy tensor $T$, the choice $\cV$ is unique up to normalization and the scaling dimension at leading order shares the same structure as the unitary case and depends on the central charge $c$, see eq. (3.23) in \cite{Behan:2025ydd},
\ba\label{eq:SR-T}
\Delta_T&|_{g=g_*} = 2 + \frac{\pi^2}{\beta_3}\frac{\Delta_{1,2}\Delta_{\chi}}{c}\,\delta +\bigO{\delta^2}\\
&=2+(1.5360301088\pm5\cdot 10^{-10})\,\delta +\bigO{\delta^2}.
\ea 

\section{Other \mathtext{$2d$}{2d} multicritical models}
\label{sec:Other}

The two long-range constructions are straightforwardly generalized to respectively the $i\vphi^{2m-1}$-theory and the universality class $\cM(2,2m+1)$, with $m\in\N$, $m\ge2$.
For comparison, we will assume as our starting assumption Sak's argument on the behaviour of fixed points in $s$, as described in and above figure \ref{fig:phase}. We will then see how this leads to potential contradictions to the necessary conditions for the correspondence between both constructions listed in the previous section.\medskip


{\bf Deformation in $i\vphi^{2m-1}$.}
Consider first the Landau-Ginzburg action at generic $m\in\N$, $m\ge2$,
\be\label{eq:MFTm-S}
S_{\text{LR}}^{(m)}=S_{\text{GFF}}[\vphi]+\frac{i\lambda_0}{(2m-1)!}\int \dd^2x\, \vphi^{2m-1}(x).
\ee
The marginality of the interaction term sets the MFT value at $\overline{s} = 2-\frac{4}{2m-1}$. Naturally, this includes the $i\vphi^3$-theory for $m=2$.
Slightly away from marginality $(2m-1)\Delta_{\vphi}=2-\eps$, or equivalently at $s=\overline{s}+\frac{2\eps}{2m-1}$, the beta function takes the general form:
\be\label{eq:MFTm-beta}
\beta(\lambda)= -\eps \lambda +B_3\lambda^3+\bigO{\lambda^5}.
\ee
Again, all even contribution vanishes due to a $\Z_2$ symmetry in the GFF action \eqref{eq:GFF-S}.
The coefficient $B_3$ is computed from the four-point function evaluated in the unperturbed GFF theory:
\be\label{eq:MFTm-B3}
B_3=\frac{(2m-1)!}{3} \pi^2\hspace{-1em} \sum_{\substack{a,\,b,\,c\,> 0\\a+b+c=2m-1}}
\prod_{k=\{a,b,c\}} \frac{1}{k!^2}\frac{\Gamma\left(1-\frac{2k}{2m-1}\right)}{\Gamma\left(\frac{2k}{2m-1}\right)},
\ee
with $B_3>0$ for all $m\ge2$.
Then, as in the $i\vphi^3$-theory, the fixed point equation admits, at leading order, three solutions: the trivial $\lambda_*=0$ solution and a complex conjugate pair of interacting fixed points:
\be 
i\lambda_* = \pm i \sqrt{\frac{\eps}{B_3}}+\bigO{\eps}.
\ee
While we could not obtain a closed form expression for \eqref{eq:MFTm-B3}, it can easily be evaluated at either finite $m$ or in the large $m$ limit:
\be
B_3\sim  \frac{\pi^3}{9\sqrt{3}}\frac{(2m-1)!^2}{(2m-2)!}\Gamma\left(\frac{2(m+1)}{3}\right)^{-6} \(\frac{\Gamma\left(\frac{1}{3}\right)}{\Gamma\left(\frac{2}{3}\right)}\)^3,
\ee
which vanishes as $m\tend \infty$.
\medskip

Then, the scaling dimension of a monomial operators $\phi^{\alpha}$, $\alpha\in\N_*$ at the fixed point takes the general form 
\be
\left.\Delta_{\vphi^{\alpha}}\right|_{\lambda=\lambda_*} = \frac{2\alpha}{2m-1}+\(\frac{\tilde{B}_3}{B_3}-\frac{\alpha}{2m-1}\)\eps +\bigO{\eps^2}.
\ee
The coefficient $\tilde{B}_3$ at generic $m$ is a complicated but finite real sum that can be evaluated at a given $m$. Its full form is given in appendix. For any $m$, it is positive for $\alpha<2m-1$ and negative otherwise. At $\alpha=2m-1$ and $\alpha=2m-2$, it simplifies to $\tilde{B}_3|_{\alpha=2m-1} = 3 B_3$ and $\tilde{B}_3|_{\alpha=2m-2} = B_3$ respectively, and
\ba
\Delta_{\vphi^{2m-1}}|_{\lambda=\lambda_*} &= 2+2\eps +\bigO{\eps^2},\\
\Delta_{\vphi^{2m-2}}|_{\lambda=\lambda_*} &= \frac{4(m-1) +\eps}{2m-1}.
\ea
The scaling dimension of the interacting operator $\vphi^{2m-1}$ is independent of $m$, while the shadow relation $\Delta_{\vphi^{2m-2}}|_{\lambda=\lambda_*}=2-\Delta_{\vphi}$ coming from the Schwinger-Dyson equations is retrieved.


\medskip

{\bf Deformation of $\cM(2,2m+1)$.}
From representation theory \cite{DiFrancesco:1997nk}, all Virasoro primary operators in the non-unitary minimal models $\cM(2,2m+1)$ are endowed with non-positive conformal weights. 
In the short-range Landau-Ginzburg formalism, the parameter order $\vphi$ is identified with the "first" Virasoro primary operator in the spectrum \cite{Katsevich:2025ojk}, which is defined as the primary with the smallest conformal dimension in absolute value,
that is the operator $\phi_{1,2}$,
\be
\Delta_{1,2} = -1 +\frac{3}{2m+3}.
\ee
Therefore, near the short-range minimal model, the long-range action, of non-trivial fixed point denoted $\text{LRMM}_{2,2m+1}(1,2)$, is built upon a linear deformation in $\phi_{1,2}$:
\be\label{eq:SRm-S}
S_{\text{LRMM}}^{'(m)}=S_{\cM(2,2m+1)}+S_{\text{GFF}}[\chi]+g_0\int \dd^2x\, \phi_{1,2}\chi,
\ee
with the associated crossover value $s^*=4-\frac{6}{2m+3}$.
In contrast to the long-range unitary minimal models $\text{LRMM}_{m,m+1}(2,2)$ \cite{Behan:2025ydd}, the crossover value does not tend to $s=2$ in the large $m$ limit, and it moves away from the mean field regime $\overline{s}$.
\medskip

\begin{figure}[h!]
    \includegraphics[width=\columnwidth]{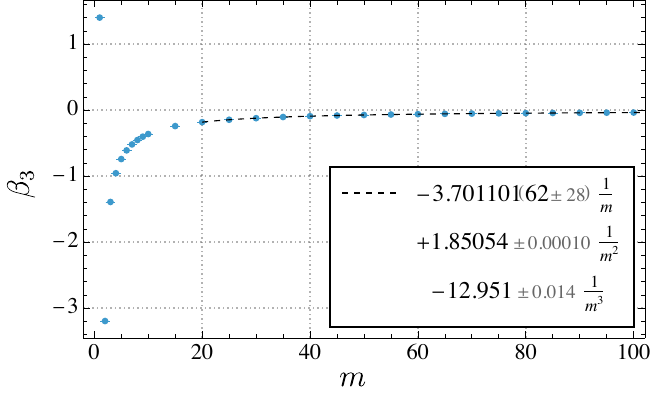}
    \caption{\label{fig:beta3}Numerical values of $\beta_3$ for $m\ge2$, fitted with a polynomial fit (up to $m^{-6}$) for $m\ge 20$. Numerical errors are dominated by the precision of the numerical integration scheme.}
\end{figure}

{In conformal perturbation theory, for $\delta = s^*-s$, the generic beta function is given by \eqref{eq:SR-beta}, with the coefficient $\beta_3$ computed now from the four-point function \eqref{eq:SR-4pts} in the unperturbed $\text{GFF}\times\cM(2,2m+1)$, with generic chiral blocks $\abs{\cF(z,\bar{z})}^2= \cF(z)\cF(\bar{z})$,}
\ba\label{eq:SRm-blocks}
&\mathcal{F}_{(1,1)}^{(1,2)(1,2)(1,k)(1,k)}(z) =\\ &(1-z)^{\frac{k-1}{2m+1}}\, {}_2 F_1 \(1-\frac{2}{2m+1},\frac{2(k-1)}{2m+1}; 2-\frac{4}{2m+1} ;z\),\\
&\mathcal{F}_{(1,3)}^{(1,2)(1,2)(1,k)(1,k)}(z) =\\ &(1-z)^{\frac{k-1}{2m+1}}z^{h_{1,3}}\, {}_2 F_1 \(\frac{2}{2m+1},\frac{2k-2m+1}{2m+1}; \frac{4}{2m+1} ;z\).
\ea
Figure \ref{fig:beta3} shows the numerical evaluation of $\beta_3$ \eqref{eq:SR-beta3} at various values of $m$.

It appears that for $\forall m>2$, $\beta_3<0$, with $\beta_3$ approaching zero from below in the extrapolated large $m$ limit. Therefore, besides the trivial fixed point, there exists a pair of interacting complex fixed points $g_*^2 = -\delta/\abs{\beta_3}$, which leads to the wrong sign in the cross-term \eqref{eq:SR-intergatingout}, and thus violates the consistency of the correspondence. Implications are raised in more details in the discussion.

Despite the lack of analytical results, careful analysis of the numerical evaluations of $\beta_3$ \eqref{eq:SR-beta3} shows that the sign's flip between $m=2$ and $m>2$ is found to originate from the $\phi_{1,3}$'s OPE channel, of associated OPE coefficient \cite{Esterlis:2016psv}:
\ba
&\left.C_{(1,2)(1,2)(1,3)}\right|_{m=2} = i\, \frac{\Gamma\(\frac{6}{5}\)}{\Gamma\(\frac{4}{5}\)}\sqrt{\frac{\Gamma\(\frac{1}{5}\)\Gamma\(\frac{2}{5}\)}{\Gamma\(\frac{3}{5}\)\Gamma\(\frac{4}{5}\)}}\in i \R,\\
&\left.C_{(1,2)(1,2)(1,3)}\right|_{m>2} = -\sqrt{\frac{2m-3}{2m-5}\frac{\Gamma\(\frac{2}{2m+1}\)}{\Gamma\(\frac{4}{2m+1}\)^2}}\times\\
&\sqrt{\frac{\Gamma\(\frac{6}{2m+1}\)\Gamma\(1-\frac{4}{2m+1}\)\Gamma\(2-\frac{4}{2m+1}\)}{\Gamma\(2-\frac{6}{2m+1}\)\Gamma\(1-\frac{2}{2m+1}\)}}\in \R.
\ea
The complexity of the IR fixed point is not reflected on the anomalous dimension of Virasoro primary operators $\phi_{1,k}$, $1\le k < m+1$, which have been verified real and negative on more than fifty operators with values of $m$ ranging from $m=3$ to $m=100$, nor on the anomalous dimension of higher broken quasiprimary currents, which have been computed up to spin $6$ for $m=3$ to $m=100$.
However, from \eqref{eq:SR-T}, in the infrared, the stress-energy tensor becomes slightly relevant and drops below the unitary bounds, 
\be
\Delta_T|_{g=g_*} = 2 + \gamma_T\,\delta +\bigO{\delta^2}, \quad \forall m>2,\, \gamma_T <0.
\ee
Thus, the long-range fixed points are unstable and do not correspond to infrared fixed points.

\section{Discussion}
\label{sec:conclusion}

Let us now conclude by summarizing the main observations of this paper.
Two types of independent long-range constructions have been studied in two dimensions: the long-range deformation of a Landau-Ginzburg formalism, the $i\vphi^{2m-1}$-theory, and of its conjectured associated minimal model $\cM(2,2m+1)$.\medskip

In the former case, nonlocality has been implemented by perturbing a GFF instead of the usual short-range Gaussian field, partially defined by its parameter $s\in\R$ dependent scaling dimension. There, for all $m\ge2$ and $s>\overline{s}$, a complex conjugate pair of non-trivial IR fixed points is identified in conformal perturbation theory as long-range models. Despite the complexity of the coupling, by \cite{Gorbenko:2018ncu}'s classification, such theories correspond to non-unitary real Euclidean QFTs.
The anomalous dimensions at leading order in conformal perturbation theory of low-lying UV composite scalar operators are shown to be real and verify the shadow relation provided by the Schwinger-Dyson equations.
However, concerning the conformality of such long-range fixed points, no proof are known beyond perturbation theory\footnote{The argument at all order in perturbation theory made in \cite{Paulos:2015jfa} only holds for $0<s<2$.} for $s\ge2$ or equivalently $\Delta_{\vphi}\le0$.\medskip

\begin{figure}[h]
   \begin{minipage}{\columnwidth}
    \centering
    \begin{tikzpicture}[glow/.style={%
    preaction={draw,line cap=round,line join=round,
    opacity=0.3,line width=4pt,#1}},glow/.default=yellow,
    transparency group]
      \tikzset{line/.style={thin}}
     
      \path[draw,-latex] (0,-1) to (5.5,-1) node [label=right:$s$,inner sep=0pt] {};

      \path[draw,-latex] (0.5,-1.2) node [label=below:{$\bar{s}=\frac{2}{3}$},inner sep=0pt] {} to (0.5,3) node [label=left:$\Delta$,inner sep=0pt] {};
      \path[glow={orange}] (0.5,-1.2) to (0.5,3);

      \path[draw,-] (4.5,-1.2) node [label=below:{$s^*=\frac{14}{5}$},inner sep=0pt] {} to (4.5,3);
      \path[glow={cyan}] (4.5,-1.2) to (4.5,3);

      \path[draw,dashed,gray] (0,0) to (5,0) node [label=right:{$\Delta=0$},inner sep=0pt] {};
      \path[draw,dashed,gray] (0,2) to (5,2)  node [label=right:{$\Delta=2$},inner sep=0pt,yshift=-1pt] {};
      \path[draw,dashed,gray] (14/6,-1.2) node [label=below:{$s=2$},inner sep=0pt] {} to (14/6,3);

      \path[draw,thick] (0.5,2/3) node [label=left:{$\vphi$},inner sep=0pt,xshift=-5pt] {} to (4.5,-4/5) node [label=right:{$\phi_{1,2}$},inner sep=0pt,yshift=3pt] {};
      \path[draw,thick] (0.5,4/3) node [label=left:{$\vphi^2$},inner sep=0pt] {} to (4.5,2.8) node [label=right:{$\chi$},inner sep=0pt] {};
      \path[draw,red] (0.5,2) to (0.75,5/2);
      \path[draw] (0.5,2) node [label=left:{$\vphi^3$},inner sep=0pt,yshift=7pt] {};
      \path[draw,red] (4.5,2) to (4.25,5/2);
      \path[draw] (4.5,2) node [label=right:{$\chi\phi_{1,2}$},inner sep=0pt,yshift=8pt] {};
      \path[draw,green!65!black] (4.5,2) to (4.25,19/8);
      \path[draw] (4.5,2) node [label=right:{$T$},inner sep=0pt,yshift=-7pt] {};

      

  \end{tikzpicture}%
    \caption{\label{fig:spectrum} Correspondence in scaling dimensions of important operators between the long-range $i\vphi^3$-theory near mean field theory (in orange) and $\text{LRMM}_{2,5}(1,2)$ near the short-range end (in blue), with the leading order in perturbation theory sketched near the relevant end.}
    \end{minipage} 
\end{figure}
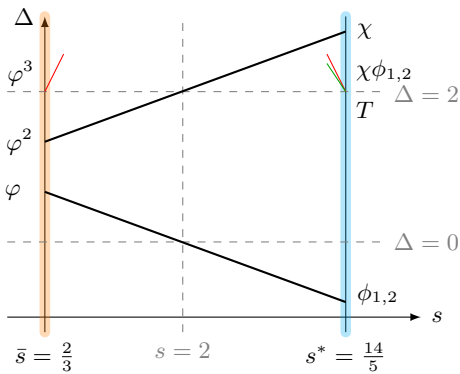

In the second construction, a separate GFF $\chi$ was coupled to the short-range $\cM(2,2m+1)$ by an interaction of the form $\int\dd^2 x\, \phi_{1,2}\chi$, only relevant under a given crossover value $s<s^*$.
At $m=2$, that is in the case of the long-range Lee-Yang model, the conformal perturbation RG flow admits a pair of real IR fixed points, consistent with the real QFT found from the $i\vphi^3$ flow.
In addition, the two Virasoro primary operators in the UV theory, $\phi_{1,2}$ and $\chi$, have protected dimensions, which again matches the prediction from the previous construction.
The correspondence is summarized in figure \ref{fig:spectrum}.

{Instead, for all $m>2$, two inconsistencies are observed:
\begin{itemize}
    \item the RG flow from the $\cM(2,2m+1)$ side admits a pair of complex fixed points;
    \item the Virasoro stress-energy tensor is slightly relevant in the IR.
\end{itemize}}
While the theory is by definition non-unitary, the imaginary fixed point induces a sign's flip in the long-range kinetic term \eqref{eq:SR-intergatingout} after integrating out the GFF $\chi$. As a consequence, the energy is unbounded from below, and the family of fixed points is inconsistent with the non-unitary yet real long-range $i\vphi^{2m-1}$ QFT.
A potential scenario around this inconsistency is that, from the point of view of the long-range $i\vphi^{2m-1}$-theories, the fixed point becomes complex at a given $2<s<s^*$. Near the short-range minimal model, this translates in the fixed point being unstable and requires the addition of the stress-energy tensor to the action, which in return would break rotational symmetry and violate spectrum continuity.
It implies that both constructions describe different fixed points, and cannot be dual to each other. 
This seems to mirror what have been observed in functional renormalization in the short-range case \cite{Benedetti:2026tpa}. There, the authors observed, in a setting where higher derivative terms were truncated, the annihilation of the short-range $i\vphi^{2m-1}$-theories fixed points one by one with non-perturbative fixed points of upper critical dimensions at $d=d_0$, the space-time dimension at which $\Delta_{\vphi}=0$, with only the $i\vphi^3$ fixed point spared.\medskip

In higher dimensions, while solvability of the long-range description near the crossover is lost, both $\overline{s}$ and $s^*$ are expected to move closer to $s=2$, up to the critical dimension\footnote{$d_c=6$ in the $i\vphi^3$ model.} $d_c$ at which they should coincide. 
However, the limit $s\tend 2$ is far from understood. 
At $s=2$, the otherwise finite wavefunction renormalization factor presents poles in $2-s$, such that perturbative results at finite $s$ cannot be straightforwardly continued to $s\tend2$, and a double perturbation theory around both $s=2$ and $d=d_c$ does not seem to describe properly the long-range fixed point. A naïve ressummation, as done in \cite{Honkonen:1990mr}, does not seem to be sufficient in the non-unitary case, and other more complicated diagrams have to be included. This again supports the assumptions that non-perturbation effects take place in the region $s>2$.
In this regard, it would be interesting to provide a better understanding of the potential long-range duality, even in the unitary case, by mapping properly the non-local conserved currents \cite{Fraser-Taliente:2026gdh} between both constructions, and non-perturbative studies using functional renormalization group or Hamiltonian truncation, would allows us to access the neighborhood of $s=2$, the long-range alter-ego of $d_0$.


\medskip
{\it Acknowledgment.}
The author thanks Connor Behan, Dario Benedetti and Thomas Pochart for useful discussions and comments.


\bibliographystyle{JHEP}
\bibliography{refs.bib}{}

\clearpage
\appendix
\onecolumngrid



\section{Near Mean Field Theory}
\label{ax:MFT}

A direct space approach to computations in perturbation theory of the Lagrangian \eqref{eq:MFTm-S} allows us to easily generalize the results to generic $m$ at leading order.
The reader is encourage to refer to the appendix A and B in \cite{Behan:2025ydd} for further technical details on either conformal perturbation theory or its use to compute direct space diagrams around a Gaussian theory.

\begin{center}
    \bf \small Beta function
\end{center}
In conformal perturbation theory, the unperturbed theory is Gaussian, such that all coefficients in the beta function \eqref{eq:MFTm-beta} are computed using Wick contractions.
Taking into account the $\Z_2$ symmetry in the field, the leading order coefficient is identified with the pole of the integral of the four-point function in the interacting field $\vphi^{2m-1}$ over a finite volume $V=2\pi R$: 
\be
B_3 = \left.\frac{2}{3!}\left(\frac{1}{(2m-1)!}\right)^3 \int_V \dd^2 x_1\dd x_2\ \la\vphi^{2m-1}(\infty)\vphi^{2m-1}(x_1)\vphi^{2m-1}(x_2)\vphi^{2m-1}(0)\ra_0\right|_{1/\eps \text{ pole}}.
\ee
The notation $|_{1/\eps}$ represents the coefficient of the $1/\eps$ pole. As mentioned in the main text, the fundamental field in a long-range theory does not get renormalized.
The diagrams contributing to the beta function are listed in table \ref{tab:MFT-betadiag} along with their integral contributions and combinatorial weights. The Feynman rules for direct space diagrams are described in appendix B of \cite{Behan:2025ydd} or in \cite{Kazakov:1984bw}.
As in dimensional regularization, no contribution of tadpoles and disconnected diagrams are taken into account.

\begin{table}[h!]
\centering
\begin{tabular}{|c|c|c|c|c|}
    \hline
     & Diagram & Conditions & Combinatorial factor& Integral\\
    \hline\hline
$\lambda_0^1$
& \begin{tikzpicture}[anchor=base, baseline,scale=0.7]
    \tikzset{line/.style={thin}}

    \node[circle,draw=black,fill=black,inner sep=0pt,minimum size=4pt,label=right:{$\infty$}] (x) at (3,0) {};
    \node[circle,draw=black,fill=black,inner sep=0pt,minimum size=4pt,label=left:{$0$}] (0) at (0,0) {};
    \draw (x) -- (0) node [midway, above] {$2m+1$};
\end{tikzpicture} & & $1$ & \\
    \hline
$\lambda_0^3$ & \begin{tikzpicture}[anchor=base, baseline=(current bounding box.center),scale=0.55]
    \tikzset{line/.style={thin}}

    \node[circle,draw=black,fill=black,inner sep=0pt,minimum size=4pt,label=right:{$x_1$}] (x1) at (4,0) {};
    \node[circle,draw=black,fill=black,inner sep=0pt,minimum size=4pt,label=left:{$0$}] (0) at (0,0) {};
    \node[circle,draw=black,fill=black,inner sep=0pt,minimum size=4pt,label=above:{$x_2$}] (x2) at (2, 2 * sqrt 3) {}; 
    \node[circle,draw=black,fill=black,inner sep=0pt,minimum size=4pt,label=below:{$\infty$}] (inf) at (2,2/3 * sqrt 3) {}; 
    \draw (0) -- (x2) node [midway, left] {$a$};
    \draw (x2) -- (x1) node [midway, right] {$b$};
    \draw (x1) -- (0) node [midway, below] {$c$};
    \draw (0) -- (inf) node [midway, above] {$d$};
    \draw (x2) -- (inf) node [midway, right] {$e$};
    \draw (x1) -- (inf) node [midway, above] {$f$};
\end{tikzpicture}&
$\begin{aligned}
        &a=f,\,b=d,\,c=e,\\
        &a+b+c=2m-1,\\
        &a,b,c\neq0
\end{aligned}$ &
$\displaystyle\sum_{\substack{a+b+c=2m-1\\ a,\,b,\,c\,\neq 0}}\frac{1}{3}\frac{(2m-1)!}{(a!b!c!)^2}$&$D_3(a,b,c)$\\
\hline
\end{tabular}
\caption{Diagrams contributing to the beta function for generic $m$.}
\label{tab:MFT-betadiag}
\end{table}


The only contributing integral does so with a simple pole, as expected from the vanishing previous orders.
\be\label{eq:MFTax-D3}
D_3(a,b,c)= \int_V \dd^2 x_1\dd^2 x_2\, \frac{1}{\abs{x_1}^{2c\Delta}}\frac{1}{\abs{x_2}^{2a\Delta}}\frac{1}{\abs{x_1-x_2}^{2b\Delta}} = \frac{\pi^2}{\eps} R^{2\eps} \prod_{k=\{a,b,c\}} \frac{\Gamma\left(1-\frac{2k}{2m-1}\right)}{\Gamma\left(\frac{2k}{2m-1}\right)} + O(\eps).
\ee

The resulting final coefficient $B_3$ is reported in \eqref{eq:MFTm-B3}. The expansion in the large $m$ limit is closely related to the long-range $\vphi^{2m-2}$ case \cite{Behan:2025ydd} and will not be repeated here.

\begin{center}
    \bf \small Anomalous dimension of monomial operators
\end{center}

The scaling dimension of the interacting field $\vphi^{2m-1}$ at the IR fixed point is directly obtained from the derivative of the beta function and independent of $m$:
\be\label{eq:MFTax-anointer}
\left.\Delta_{\vphi^{2m-1}}\right|_{\lambda=\lambda_*} = 2+\beta'(\lambda_*) = 2 +2\eps +\bigO{\eps^2}.
\ee
The anomalous dimensions of other monomials operators, $\vphi^\alpha$, $\alpha\in \N$, are obtained by renormalizing the two-point function. Defining, the bare operator $\vphi^\alpha = Z_{\alpha}\vphi^\alpha_r$, the anomalous dimension is then given as:
\be
\gamma_\alpha(\lambda) = -\frac{1}{Z_\alpha}R\frac{\dd Z_\alpha}{\dd R}, \quad \Delta_{\vphi^\alpha} = \alpha\Delta_{\vphi} +\gamma_\alpha(\lambda_*).
\ee

In conformal perturbation theory using the $\overline{\text{MS}}$ scheme, the leading order in the anomalous dimension is given by
\be
\gamma_{\alpha}(\lambda) = \tilde{B}_3 \lambda^2 +\bigO{\lambda^4},
\ee
with the coefficient $\tilde{B}_3$ once again controlled by the pole of the integral of the four-point function of the interacting field $\vphi^{2m-1}$ with the operator considered $\vphi^\alpha$:
\ba
    \tilde{B}_3 &= \left.\left(\frac{1}{(2m-1)!}\right)^2 \int_V \dd^2 x_1\dd x_2\, \la\phi^{\alpha}(\infty)\phi^{\alpha}(0)\phi^{2m-1}(x_1)\phi^{2m-1}(x_2)\ra_0\right|_{1/\eps \text{ pole}}.\\
\ea
The list of diagrams contributing are given in table \ref{tab:MFT-anodiag}, where we defined
\be
\tilde{S}_{d}^{a,b,c} = \(\frac{\alpha!}{a!c!}\)^2\frac{1}{b!d!}.
\ee
Tadpoles and disconnected diagrams do not contribute to the UV divergence.

\begin{table}[h!]
\begin{tabular}{|c|c|l|c|c|}
    \hline
     & Diagram & Conditions & Combinatorial factor & Integral\\
    \hline\hline
$\lambda_0^0$
& \begin{tikzpicture}[anchor=base, baseline,scale=0.7]
    \tikzset{line/.style={thin}}

    \node[circle,draw=black,fill=black,inner sep=0pt,minimum size=4pt,label=right:{$\infty$}] (x) at (3,0) {};
    \node[circle,draw=black,fill=black,inner sep=0pt,minimum size=4pt,label=left:{$0$}] (0) at (0,0) {};
    \draw (x) -- (0) node [midway, above] {$\alpha$};
\end{tikzpicture} & & $\alpha!$ & \\
\hline
$\lambda_0^2$ & \begin{tikzpicture}[anchor=base, baseline=(current bounding box.center),scale=0.55]
    \tikzset{line/.style={thin}}

    \node[circle,draw=black,fill=black,inner sep=0pt,minimum size=4pt,label=right:{$x_1$}] (x1) at (4,0) {};
    \node[circle,draw=black,fill=black,inner sep=0pt,minimum size=4pt,label=left:{$0$}] (0) at (0,0) {};
    \node[circle,draw=black,fill=black,inner sep=0pt,minimum size=4pt,label=above:{$x_2$}] (x2) at (2, 2 * sqrt 3) {}; 
    \node[circle,draw=black,fill=black,inner sep=0pt,minimum size=4pt,label=below:{$\infty$}] (inf) at (2,2/3 * sqrt 3) {}; 
    \draw (0) -- (x2) node [midway, left] {$a$};
    \draw (x2) -- (x1) node [midway, right] {$b$};
    \draw (x1) -- (0) node [midway, below] {$c$};
    \draw (0) -- (inf) node [midway, above] {$d$};
    \draw (x2) -- (inf) node [midway, right] {$c$};
    \draw (x1) -- (inf) node [midway, above] {$a$};
\end{tikzpicture}
& \small$\begin{aligned}
        &0<a<\min\left(\alpha,2m-1\right)-1\\
        &\max\left(0,2m-1-\alpha\right)<b<2m-1-a\\
        &0<c=2m-1-a-b\\
        &0 < d = b+\alpha-2m+1\\
    \end{aligned}$
&\makecell{$\begin{aligned}\displaystyle\sum_{a=1}^{\min(\alpha,2m-1)-2}\hspace{-12pt}&\sum_{b=\max(0,2m-1-\alpha)+1}^{2m-2-a}\\&\vphantom{\int^{a^{a^a}}}\tilde{S}^{a,b,2m-1-a-b}_{b+\alpha-2m+1}\end{aligned}$
} & $D_3(a,b,c)$ \\
\cline{2-5}
 & \begin{tikzpicture}[anchor=base, baseline=(current bounding box.center),scale=0.55]
    \tikzset{line/.style={thin}}

    \node[circle,draw=black,fill=black,inner sep=0pt,minimum size=4pt,label=right:{$x_1$}] (x1) at (4,0) {};
    \node[circle,draw=black,fill=black,inner sep=0pt,minimum size=4pt,label=left:{$0$}] (0) at (0,0) {};
    \node[circle,draw=black,fill=black,inner sep=0pt,minimum size=4pt,label=above:{$x_2$}] (x2) at (2, 2 * sqrt 3) {}; 
    \node[circle,draw=black,fill=black,inner sep=0pt,minimum size=4pt,label=below:{$\infty$}] (inf) at (2,2/3 * sqrt 3) {}; 
    \draw (0) -- (x2) node [midway, left] {$a$};
    \draw (x2) -- (x1) node [midway, right] {$b$};
    \draw (x1) -- (0) node [midway, below] {$c$};
    \draw[dashed] (0) -- (inf);
    \draw (x2) -- (inf) node [midway, right] {$c$};
    \draw (x1) -- (inf) node [midway, above] {$a$};
\end{tikzpicture}
& \small$\begin{aligned}
        &\alpha<2m-1\\
        &0<a<\alpha\\
        &0<b=2m-1-\alpha\\
        &0< c = \alpha-a\\
    \end{aligned}$
&$\begin{aligned}\Theta(2m-1-\alpha)\times\\\displaystyle\sum_{a=1}^{\alpha-1}\tilde{S}^{a,2m-1-\alpha,\alpha-a}_{0}\end{aligned}$ & $D_3(a,b,c)$ \\
\hline
\end{tabular}
\caption{Diagrams contributing to the anomalous dimension of $\phi^\alpha$, $\alpha\in \N$ for generic $m$. $\Theta$ denotes the Heaviside function.}
\label{tab:MFT-anodiag}
\end{table}

Putting all contributions together gives:
\ba\label{eq:MFTax-B3ano}
\tilde{B}_3 =& \frac{\pi^2}{\alpha!}\left[\sum_{a=1}^{\min(\alpha,2m-1)-2}\sum_{b=\max(0,2m-1-\alpha)+1}^{2m-2-a}\tilde{S}^{a,b,2m-1-a-b}_{b+\alpha-2m+1}\prod_{k=\{a,b,2m-1-a-b\}} \frac{\Gamma\left(1-\frac{2k}{2m-1}\right)}{\Gamma\left(\frac{2k}{2m-1}\right)}\right.\\
&+\left. \Theta(2m-1-\alpha)\sum_{a=1}^{\alpha-1}\tilde{S}^{a,2m-1-\alpha,\alpha-a}_{0}\prod_{k=\{a,2m-1-\alpha,\alpha-a\}} \frac{\Gamma\left(1-\frac{2k}{2m-1}\right)}{\Gamma\left(\frac{2k}{2m-1}\right)}\right].
\ea

While no closed form are known, the coefficient is easily evaluated at finite $m$ using Mathematica, and the relations \eqref{eq:MFTax-anointer} and $\Delta_{\vphi^{2m-2}}|_{\lambda=\lambda_*} = 2-\Delta_{\vphi}$ are retrieved.

\section{Non-unitary minimal models}
\label{ax:MM}

\begin{center}
    \bf \small Conventions
\end{center}

We consider the $\cM_{2,2m+1}$ class of diagonal minimal models, with negative central charge:
\be
c = -2(3m-5)-\frac{12}{2m+1}, \quad m\in \N,\ m>1.
\ee
Holomorphic Virasoro primaries $\phi_{1,k}(z)$ are labeled by a positive integer $1\le k\le 2m$, of non-positive scaling dimension \be
\Delta_{1,k} = 2 h_{1,k}, \quad h_{1,k}= -\frac{(k-1)(2m-k)}{4m+2}.
\ee
The operator $\phi_{1,1}\equiv \id$ is identified with the identity, and other operators are degenerate in pairs $(1,k)\simeq (1,2m+1-k)$. We assume their two-point functions to be unit-normalized. The generic fusion rules form a closed sector and are given by
\be
\phi_{1,k}\times \phi_{1,k'} = \sum_{\substack{k''=\abs{k-k'}+1\\k+k'+k'' \text{ odd}}}^{\min(k+k'-1,4m+1-k-k')} \phi_{1,k''}.
\ee
In particular, $\phi_{1,2}\times \phi_{1,2} = \id + \phi_{1,3}$.

\begin{center}
    \bf \small Four-point function
\end{center}

Let's consider the correlator:
\ba
\corr{\phi_{1,k}&(z_1,\zcc_1)\phi_{1,2}(z_2,\zcc_2)\phi_{1,2}(z_3,\zcc_3)\phi_{1,k}(z_4,\zcc_4)}
=\\&\(\frac{\abs{z_{14}}}{\abs{z_{13}}}\)^{\Delta_{1,2}-\Delta_{1,k}}\(\frac{\abs{z_{24}}}{\abs{z_{14}}}\)^{\Delta_{1,k}-\Delta_{1,2}}\sum_{l={-1,1}} (C^{(1,k+l)}_{(1,2)(1,2)})^2\frac{\mathcal{F}_{(1,k+l)}(z)\mathcal{F}_{(1,k+l)}(\zcc)}{\(\abs{z_{12}}\abs{z_{34}}\)^{\Delta_{1,k}-\Delta_{1,2}}},
\ea
with $z$, $\zcc$ the cross-ratio $z = \frac{z_{12}z_{34}}{z_{13}z_{24}}$, and $z_{ij}=z_i-z_j$.
The holomorphic blocks $\mathcal{F}_k(z)$, are solution of the BPZ equation for the holomorphic part:
\begin{align}
	\mathcal{D}_2^{(z_2)}\la \phi_{1,k}(z_1)\phi_{1,2}(z_2)\phi_{1,2}(z_3)\phi_{1,k}(z_4)\ra=0\,,
\end{align}
where $\mathcal{D}_2^{(z_2)}$,  acting on $z_2$,  is the differential operator:
\be
	\mathcal{D}_2^{(\cdot)}= {\mathcal L}_{-2}^{(\cdot)}+\frac{m-1}{2 h_{1,2}}({\mathcal L}_{-1}^{(\cdot)})^2\,,\quad
    \mathcal{L}_{-k}^{(w)}\equiv\sum_{i=1}^{3}\left(\frac{ (k-1)h_i}{(z_i-w)^k}-\frac{1}{(z_i-w)^{k-1}}\partial_i\right)\,,\quad {\cal L}_{-1}^{(w)}=\partial_w\,.
\ee
The above equation corresponds to a second order differential equation, with thus two general solutions corresponding each to a different block $\mathcal{F}_k(z)$ and are identified by imposing slightly different boundary conditions for different value of $k$.
In practice, we will choose to impose that
\be
\mathcal{F}_k(z) = z^{h_k}\(1 + \frac{h_k}{2}\, z +\bigO{z^2}\).
\ee
Then, the solutions at generic $m$ and $r$ are given by:
\ba
&\mathcal{F}_{(1,k-1)}^{(1,k)(1,2)(1,2)(1,k)}(z) =(1-z)^{\frac{1}{2m+1}}z^{h_{1,k-1}}\, {}_2 F_1 \(\frac{2}{2m+1},\frac{2m+3-2k}{2m+1}, 2-\frac{2k}{2m+1} ,z\),\\
&\mathcal{F}_{(1,k+1)}^{(1,k)(1,2)(1,2)(1,k)}(z) =(1-z)^{\frac{1}{2m+1}}z^{h_{1,k+1}}\, {}_2 F_1 \(\frac{2}{2m+1},\frac{2k-2m-3}{2m+1}, \frac{2k}{2m+1} ,z\).
\ea

The holomorphic blocks \eqref{eq:SRm-blocks} are obtained in a similar fashion with now,
\ba
\corr{\phi_{1,2}&(z_1,\zcc_1)\phi_{1,2}(z_2,\zcc_2)\phi_{1,k}(z_3,\zcc_3)\phi_{1,k}(z_4,\zcc_4)}
=\abs{z_{12}}^{-\Delta_{1,2}}\abs{z_{34}}^{-\Delta_{1,k}}\(\mathcal{F}_{(1,1)}^{(1,2)(1,2)(1,k)(1,k)}(z)\mathcal{F}_{(1,1)}^{(1,2)(1,2)(1,k)(1,k)}(\zcc)\right.\\&\left.+C_{(1,2)(1,2)(1,3)}C_{(1,k)(1,k)(1,3)}\mathcal{F}_{(1,3)}^{(1,2)(1,2)(1,k)(1,k)}(z)\mathcal{F}_{(1,3)}^{(1,2)(1,2)(1,k)(1,k)}(\zcc)\).
\ea
OPE coefficient are computed via the Coulomb gas formalism \cite{Esterlis:2016psv}.


\end{document}